\documentclass[aps,pra,twocolumn,showpacs,floatfix]{revtex4}

\usepackage{bm,bbm,amsmath, amssymb,amsbsy,
amsthm,graphicx,subfigure,color,txfonts,multirow}
\usepackage{amsmath}
\usepackage{amsfonts}
\usepackage{array}
\usepackage[utf8]{inputenc}
   \newcommand{\mc}[1]{\mathcal{#1}}

\usepackage[framemethod=tikz]{mdframed}
\usepackage[colorlinks]{hyperref}
\hypersetup{
	bookmarksnumbered,
	pdfstartview={FitH},
	citecolor={blue}, 
	linkcolor={red},
	urlcolor={black},
	pdfpagemode={UseOutlines}
}

 \newcommand{\tr}[1]{\text{Tr}}
\newcommand{\ket}[1]{|#1\rangle}
\newcommand{\bra}[1]{\langle#1|}

\begin{document}

\title{Witnessing multipartite entanglement by
detecting asymmetry}

\author{Davide Girolami}
\email{davegirolami@gmail.com}
\author{Benjamin Yadin}
\affiliation{$\hbox{Clarendon Laboratory,
Department of Physics, University of Oxford,
Parks Road, Oxford OX1 3PU, United Kingdom}$
\\}

\begin{abstract}
The characterization of quantum coherence in
the context of quantum information theory and
its interplay with quantum correlations is
currently subject of intense study.
Coherence in an Hamiltonian
eigenbasis yields asymmetry, the ability of
a quantum system to break a dynamical symmetry
generated by the Hamiltonian. We here propose
an experimental strategy to witness multipartite entanglement in many-body systems by evaluating the asymmetry with respect
to an additive Hamiltonian. We test our scheme
by simulating asymmetry and entanglement
detection in a three-qubit GHZ-diagonal state.
 \end{abstract}
 
\date{\today}


  \maketitle

\section{Introduction}
Quantum Information Theory provides important
insights on the foundations of Quantum
Mechanics, as well as its technological
applications. The framework of resource
theories characterizes the quantum laws as
constraints, and the properties of quantum
systems as resources for information
processing \cite{opp}. In this context, the
degree of coherent superposition of a 
state $\sum_i c_i \ket{i}\bra{i}, \sum_i
|c_i|^2=1$, i.e. coherence (we omit the
quantum label, from now on) in a reference
basis $\{i\}$, is a resource. The crucial
question is to determine how to obtain a
computational advantage powered by coherence
\cite{ger2,china,china2,luo2,heng,plenio,me,blind,lqu,mehdi,luo,superreview,gour,newmar,herbut,aberg,luocri,speknat}.
%Several theoretical quantifiers of coherence
The coherence of a finite-dimensional quantum state
$\rho$ has been defined as its distinguishability from the sets
of states which are diagonal in a given basis
\cite{herbut,plenio,ger2,china,china2,heng}.
%We relate speed to observable quantities byadopting a geometricviewpoint. The dynamics ofa finite dimensional quantumsystem can bevisualized as a parametrized path t; t 2 R,inthe space of accessible states (Fig.1). Theparameter t representsthe evolution time.Aquantum process is modelled as a geometricpath t ,where the coordinate t represents thecomputation time. The speed
%in an interval  is defined as the averagerate of change of the systemstate along thepath from 0 to . For periodic motions, t =0; t =kT; k 2 Z; the quantity is informativeif taken over an intervalwhere the coordinatet represents the computation time. The speedinan interval  is defined as the average rateof change of the systemstate along the pathfrom 0 to .
 Yet, to date, there is no operational interpretation for such definition of coherence.
A concurrent body of work  has linked the coherence of $\rho$ in a basis $\{h\}$  to the degree of uncertainty in a  measurement of
  an observable $H=\sum_h h \ket{h}\bra{h}$ on $\rho$. Such genuinely quantum uncertainty
has been proven to have an operational interpretation, corresponding to the sensitivity of the state
to a phase shift generated by $H$
\cite{superreview,newmar,mehdi,gour,speknat,lqu,blind,me,luo,luocri,luo2,aberg}. From a physics perspective, coherence
here underpins  {\it
$U(1)$-asymmetry}. The asymmetry of a quantum
system quantifies its ability to be a
reference frame under a phase superselection
rule, where $H$ is the observable whose
coherent superpositions are prohibited (e.g.
electric charge, energy).  In other words, asymmetry is the geometric property of a quantum system which makes it able to break a symmetry generated by an Hamiltonian $H$.

Further studies bridged the gap between
these recent theoretical findings and  the 
experimental implementation of quantum information processing, by providing a strategy to measure
the   asymmetry of an arbitrary
quantum state in the laboratory with the
current technology \cite{me} (for coherence
witnesses, see \cite{agata,noriwit,felix}).
These results paved the way for investigating the link between
  coherence and
quantum properties of multipartite systems. In
particular, the relationship between coherence
and quantum correlations has been explored
\cite{ger1,ger2,lqu,blind,china,luo2,heng,alta}.
\\
In this work, we show how detecting asymmetry
in states of multipartite qubit systems allows
an experimentalist to verify entanglement with
limited resources. Entanglement is a crucial
property for quantum information processing
\cite{horo}, e.g. providing speed-up in
communication and metrology protocols
\cite{jozsa,metrorev}. Yet, it is hard to be
quantified in both theoretical and
experimental practice
\cite{toth3,morimae,huber}. On this purpose,
we here introduce an experimentally friendly
witness of multipartite entanglement in terms
of the asymmetry with respect to an additive
Hamiltonian. \\
The structure of the paper is the following.
In Sec.~\ref{sect2a}, we recall that the
quantum Fisher information, a measure of
sensitivity of a state to phase shifts
employed in quantum metrology
\cite{helstrom,toth,metrorev,speed}, is an asymmetry quantifier. It is
possible to identify a lower bound of it in
terms of traces of density matrix powers. We calculate how much the experimentally reconstructed bound deviates from the theoretical quantity (Sec.~\ref{sect2b}). Also, we  express the lower bound for one, two and three-qubit states in terms of finite phase shifts generated by spin observables.
These quantities can be evaluated by single qubit interferometry
\cite{filip,ekertdir,paz,brun,dariano}, as well
as local projective measurement schemes
\cite{me,pasca,jeong2,geodiscordchina,mintert,mintert2,kus},
without performing full state reconstruction.
In Sec.~\ref{sect3a}, we show that the
asymmetry lower bound witnesses genuinely multipartite entanglement when
measured with respect to an additive multipartite
Hamiltonian.
We complete the study with a demonstrative
example (Sec.~\ref{sect3b}). We simulate the
evaluation of asymmetry and entanglement in a
GHZ-diagonal state by a seven-qubit quantum
information processor. We draw our conclusions
in Sec.~\ref{sect4}.

\section{Measuring asymmetry}

\subsection{Theoretically consistent measure of asymmetry}\label{sect2a}

\subsubsection*{Quantum Fisher information}
In the resource theory of  
asymmetry
\cite{gour,superreview,newmar,mehdi,speknat},
the consumable resource is any system whose
state is not commuting with a fixed, bounded
observable $H$ with spectral decomposition
$H=\sum_h h \ket h\bra h$. A system in the
incoherent state $\rho_H$ such that
$\rho_H=\sum_h c_h \ket{h}\bra h,
[\rho_H,H]=0,$ is free, in the sense that it
can be arbitrarily added or discarded in a
quantum protocol without affecting the
available asymmetry. Free states are
invariant under phase rotations generated by
$H$: $ e^{-iH\theta} \rho_H e^{i H\theta} =
\rho_H, \quad \forall \theta \in \mathbb{R}$.
The free operations are the CPTP
(completely-positive trace-preserving) maps $\mc{E}_H$ which
cannot increase the amount of asymmetry in a
state.
They are identified by maps commuting with the unitary evolution generated by the observable under scrutiny, $e^{-i H\theta} \mc{E}_H(\rho)
e^{i H\theta} = \mc{E}_H(e^{-i H \theta}\rho
e^{i H\theta}), \quad \forall \rho, \theta$. 
Their explicit form is studied in
Ref.~\cite{gour}.
Several quantifiers of asymmetry have been
proposed \cite{gour,speknat,me}. Here we adopt the viewpoint of asymmetry as a measure of the
state usefulness in a phase estimation
scenario. The symmetric logarithmic derivative quantum Fisher
information is indeed a measure of asymmetry \cite{speed,speknat}. Let us
recall its definition. Given the spectral
decomposition of a probe state $\rho = \sum_i
\lambda_i |i\rangle\langle i|, \sum_i
\lambda_i =1,$ and an observable $H$, the
quantum Fisher information ${\cal F}_H(\rho) =
2 \sum_{i,j}
\frac{(\lambda_i-\lambda_j)^2}{\lambda_i+\lambda_j}\,H_{ij}^2,
H_{ij}= |\langle i|H|j\rangle|,$ quantifies
the sensitivity of the probe to a phase shift
$U_H(\theta)=e^{-i H \theta}$ generated by  $H$, under the assumption that the
state changes smoothly \cite{helstrom}.
The quantum Fisher information is (four times)
the convex roof of the variance, ${\cal
V}_H(\ket{\psi}) := 4 \left(
\langle\psi|H^2|\psi\rangle -
\langle\psi|H|\psi\rangle^2 \right)$, meaning
that $ {\cal F}_H(\rho) = \inf_{\{ p_i,
\ket{\psi_i} \}} \, \sum_i p_i {\cal
V}_H(\ket{\psi_i})$, where the infimum is
taken over all the convex decompositions $\rho
= \sum_i p_i |\psi_i\rangle\langle\psi_i|$
such that the $\{p_i\}$ form a probability
distribution \cite{yu,toth2}. Moreover, a
decomposition saturating the equality always
exists. This  property implies convexity,  
${\cal F}_H(p\rho + (1-p)\sigma) \leq p\,
{\cal F}_H(\rho)+ (1-p){\cal F}_H(\sigma)$. The quantum Fisher information is equal to the variance for pure states, $
{\cal F}_H(|\psi\rangle\langle\psi|) = {\cal
V}_H(|\psi\rangle)$.\\
We recall what implies that the quantum Fisher information is a reliable
measure of asymmetry. It satisfies the
following criteria:\\
i) It vanishes if and only if the state is
incoherent. Since the quantum Fisher
information is convex, for any incoherent
state one has ${\cal F}_H(\rho_H)={\cal
F}_H(\sum_h c_h \ket h\langle h|)\leq \sum_h
c_h {\cal F}_H(\ket h\langle h|)=0$. Also, we
observe that ${\cal F}_H(\rho)=0
\Leftrightarrow [\rho,H]=0,$ and $ [\rho,H]=0
\Leftrightarrow [\rho, U_H(\theta)]=0, \forall
\theta$, which is a condition satisfied if and
only if the state is incoherent.\\
ii) It cannot increase under symmetric
operations. Given
$H_{AB}=H_A\otimes I_B+I_A\otimes
H_B$, by Theorem II.1 of Ref.~\cite{mehdi}, any
incoherent map $\mc{E}_{H_A}$ admits a
Stinespring dilation ${\cal
E}_{H_A}(\rho_A)=\text{Tr}_B[V^{H}_{AB}(\rho_A\otimes\tau_B)V^{H\dagger}_{AB}]$, where
$V^H_{AB}$ is a symmetric unitary with respect to $H_{AB}$, and
$[\tau_B,H_B]=0$. In other words, any
symmetric map can be represented by the  
unitary, symmetric evolution of the system of
interest and an ancilla in an incoherent
state. One then obtains ${\cal
F}_{H_A}(\rho_A)={\cal
F}_{H_{AB}}(\rho_A\otimes\tau_B)={\cal
F}_{H_{AB}}(V^{H\dagger}_{AB}(\rho_A\otimes\tau_B)V^{H}_{AB})\geq
{\cal
F}_{H_A}(\text{Tr}_B[V^{H\dagger}_{AB}(\rho_A\otimes\tau_B)V^{H}_{AB}])={\cal
F}_{H_A}({\cal E}_{H_A}(\rho_A))$. \\
The proof can be extended to any quantum Fisher information $
 {\cal I}^{f}_H(\rho)=\sum_{i,j} \frac{(\lambda_i-\lambda_j)^2}{\lambda_j f(\lambda_i/\lambda_j)} |\langle i|H| j\rangle|^2,  
$ where each of the real-valued functions
 $f$ identifies a quantization of the classical   Fisher information which preserves contractivity under noisy operations, being ${\cal F}_H(\rho)= {\cal I}^{F}_H(\rho), F(x)=(1+x)/2, x\in\mathbb{R}$ \cite{petzmono}. The quantum Fisher informations are topologically equivalent, being connected by the chain $2 f(0){\cal I}^f_H(\rho)\leq {\cal F}_H(\rho)\leq  {\cal I}_H^f(\rho),\forall f, \rho, H$ \cite{gibilisco}.
 Also, the property ii) can be generalized to show that any quantum Fisher information is an ensemble monotone, i.e. it does not increase on average under symmetric operations, $
{\cal I}_H^f(\rho)\geq\sum_{\mu}p_{\mu} {\cal I}^f_H({\cal E}_{\mu}(\rho)),
 \forall \{p_{\mu}, {\cal E}_{\mu}\}:\sum_{\mu}p_{\mu}=1, [{\cal E}_{\mu},U_H(\theta)]=0, \forall f   
$ \cite{speed}. 

\subsubsection*{Asymmetry lower bound}
Picking the Fisher information as a measure of
asymmetry is useful for experimental purposes.
Coherence is not a linear property of a system, so it cannot be directly related to a quantum operator \cite{diogo}. Also, the quantum
Fisher information is usually hard to be
computed. Yet, it is possible to build up an
observable quantity which provides a nontrivial
lower bound:\\
\begin{eqnarray}\label{lowbound}
{\cal O}_H(\rho)&\leq& {\cal F}_H(\rho), \\
{\cal O}_H(\rho)&=&-2\text{Tr}[[\rho,H]^2]=4
\text{Tr}[\rho^2 H^2 - \rho H \rho
H]\nonumber.
\end{eqnarray} 
As observed in Ref.~\cite{speed}, one has ${\cal
O}_H(\rho)=2\sum_{i\neq j}(\lambda_i
-\lambda_j)^2 H_{ij}^2$. Since
$\lambda_i+\lambda_j \leq 1, \forall i,j $, by
recalling the expression of the quantum Fisher
information, the lower bound holds.   For pure states, one has $ {\cal O}_H(\rho)= {\cal
F}_H(\rho) =4{\cal
V}_H(\rho)$. The lower bound reliably detects asymmetry, as ${\cal
O}_H(\rho)=0 \Leftrightarrow {\cal
F}_H(\rho)=0$. \\
One may wonder if the quantity ${\cal O}_H(\rho)$ itself is a
consistent measure of asymmetry.
For pure states, the lower bound equals the
quantum Fisher information, so the answer is
positive in such a case. Unfortunately, this
does not hold for mixed states. We can see
that with a simple example. Given a bipartite
state $\rho_{AB}=\rho_A \otimes \rho_B$, let
us suppose to measure the asymmetry of the
marginal state $\rho_A$ as the uncertainty
measuring $H_A$. One obtains ${\cal O}_{H_A}
(\rho_{AB})= {\cal
O}_{H_A}(\rho_A)\text{Tr}[\rho_B^2]$. Then,
discarding the subsystem $\rho_B$ would
increase the asymmetry of the state $\rho_A$,
which is manifestly undesirable. One may
normalize the quantity by employing ${\cal
O}_H(\rho)/\text{Tr}[\rho^2]$ as a measure of
asymmetry, yet there would still be a problem.
Note that the bound is written (modulo a
constant) as an Hilbert-Schmidt norm in the zero shift limit, $ {\cal
O}_H(\rho)=2\lim_{\theta\rightarrow 0}||U_H(\theta)\rho U_H(\theta)^{\dagger}-\rho||_{2}^2/(\theta^2)$. This
norm is notoriously not contractive under
quantum operations \cite{schirmer}. Not
surprisingly, this property also makes
measures of quantum correlations based on this
norm generally unreliable \cite{piani,tufo}.
Hence, the lower bound, while being not a
full-fledged measure, can replace the quantum Fisher
information in scenarios
where some restriction is posed, e.g. for unitary evolutions of  systems which are guaranteed to be closed.\\

\subsection{Experimental observability of
the asymmetry bound}\label{sect2b} 

\subsubsection*{Experimental scheme}
\begin{figure}[t]
\includegraphics[height=7cm,width=7cm]{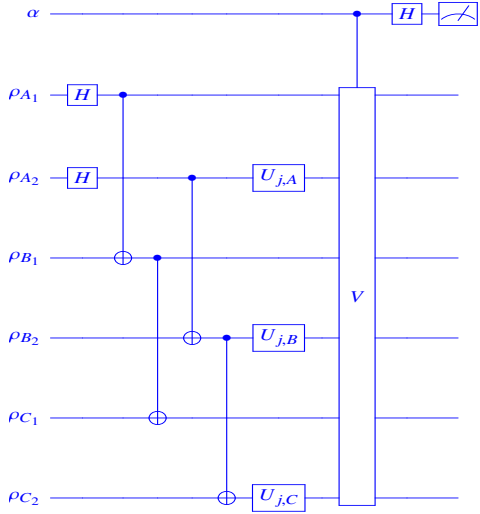}
\caption{Overlap detection. Two copies
$\rho^p_{A_1B_1C_1}, \rho^p_{A_2B_2C_2}$ of a
GHZ-diagonal state are prepared in the state $\rho_i=1/2
(I_2+ p \sigma_z),\forall i$. An Hadamard gate $H$ is applied to 
the qubits $A_i, i=1,2$, followed by two CNOT
gates on each copy. Then, one evaluates the
purity and the overlap terms related to
the observables $J_{3,x(y,z)}$, by applying the unitary
transformations
$U_{J_3}(\theta)=U_{j,A}(\theta)\otimes
U_{j,B}(\theta)\otimes U_{j,C}(\theta),$
and measuring the ancilla polarisation by
means of an interferometric scheme. This
consists of an ancilla in the initial state
$\alpha = 1/\sqrt{2}(\ket{0}+\ket{1})$
interacting with the two state copies by a
controlled-$V$ gate, being $V$ the swap
operator. A second Hadamard gate $H$ is finally
applied to the ancilla. The mean value of the
ancilla polarisation at the output is
$\langle\sigma_z\rangle_{\alpha^{\text{out}}}=
\text{Tr}[V \rho\otimes U_{J_3}(\theta)\rho
U^{\dagger}_{J_3}(\theta)]= \text{Tr}[\rho
U_{J_3}(\theta)\rho
U^{\dagger}_{J_3}(\theta)]$, which determines
the asymmetry lower bound.}
\label{complete}
\end{figure}

As shown in Ref.~\cite{me}, the asymmetry lower
bound  is a function of mean
values of self-adjoint operators. By applying
the Taylor expansion about $\theta=\theta_0$,
one has $\text{Tr}[\rho U_H(\theta) \rho
U_H^{\dagger}(\theta)]=\text{Tr}[\rho
U_H(\theta_0) \rho U_H^{\dagger}(\theta_0)]-
(\text{Tr}[\rho^2 H^2]-\text{Tr}[\rho H\rho
H])(\theta-\theta_0)^2 +
O((\theta-\theta_0)^3)$, and then ${\cal
O}_H(\rho)\sim\frac{4}{(\theta-\theta_0)^2}(\text{Tr}[\rho
U_H(\theta_0) \rho U_H^{\dagger}(\theta_0)]
-\text{Tr}[\rho U_H(\theta) \rho
U_H^{\dagger}(\theta)]), \theta \rightarrow
\theta_0$. By setting $\theta_0=0$, an approximation  in terms of finite phase shifts, with error $O(\theta^2)$, is given
by ${\cal O}_H^{\text{ap}}(\rho)\pm
\Delta{\cal O}_H^{\text{ap}}(\rho)$, with
\begin{eqnarray}\label{app}
{\cal
O}_H^{\text{ap}}(\rho)&=&4\frac{\text{Tr}[\rho^2]-\text{Tr}[\rho
U_H(\theta) \rho
U_H^{\dagger}(\theta)]}{\theta^2},\\
\Delta{\cal O}_H^{\text{ap}}(\rho)&=&
\underbrace{|d {{\cal O}_H^{\text{ap}}(\rho)}/
d{\theta}|_{\theta=0}}_{=0} \theta + 1/2 |d^2
{{\cal O}_H^{\text{ap}}(\rho)}/
d{\theta}^2|_{\theta=0}\theta^2.\nonumber
\end{eqnarray} 
One may note that even the approximated quantity is a lower bound (but less tight) to the quantum Fisher information,  $ {\cal
O}_H^{\text{ap}}(\rho)\leq {\cal O}_H(\rho), \forall \rho, H, \theta$ \cite{speed}. 
Therefore, to quantify the lower bound to the
asymmetry of the state, we need to evaluate its
purity and the overlap with a second copy of
the state after a rotation has been applied. 
They are obtained by estimating the
mean value of the swap operator
$V=\sum_{ij}\ket{ij}\bra{ji}$ in two copies of the system, $\rho_{1,2}\equiv \rho$:
$\text{Tr}[\rho^2]=\text{Tr}\Big[V(\rho\otimes\rho)\Big]$,
while the overlap is given by
$\text{Tr}\Big[\rho U_H \rho
{U_H^{\dagger}}\Big]=\text{Tr}\Big[V\Big(\rho\otimes
U_{H}\rho{U_{H}^{\dagger}}\Big)\Big].$ Such quantities can be directly measured by implementing an interferometric configuration
\cite{paz,brun,dariano,ekertdir,filip,me,jeong2,rugg}. In fact, the method has general validity regardless
the system state and the self-adjoint operator
to be measured \cite{paz}. Alternatively, for
the relevant case of N-qubit systems, it is
possible to extract purity and overlap by
local Bell measurements, a routine measurement scheme in
optical setups
\cite{mintert,mintert2,me,pasca,jeong2,geodiscordchina,kus,speed}.
Thus, for systems of arbitrary dimension, the lower bound $O_H(\rho)$ can be extracted by the statistics of a limited
number of detections, bypassing full state reconstruction.  
\begin{table*}[t]
    {\footnotesize
\begin{tabular}{|c|c|c|c|}
\hline
 $J_3 $&$J_{3,x}$& $J_{3,y}$  &$ J_{3,z}$\\
\hline
${\cal F}_{J_3}\left(\rho^p_{ABC}\right)$& $\frac{2 p^2
\left(p^2+2\right)}{p^2+1}$ & $\frac{(-2 p^8+p^6+18 p^4+7 p^2)}{3 p^4+4
p^2+1}$
&   $ (2 p^4+4 p^3+3p^2)$ \\
\hline
${\cal O}_{J_3}\left(\rho^p_{ABC}\right)$& $ (2p^2 + 3 p^4 +
p^6)/2$& $ (p^6+4
p^4+7p^2)/4$
& $ (3 p^6+8 p^5+14 p^4+8
p^3+3p^2)/4$ \\
\hline
${\cal F}_{J_3}\left(\rho^p_{ABC}\right)>3,5$ &$p>1$& $p>1$
 &$p > 0.674, 0.813$   \\
 \hline
$ {\cal O}_{J_3}\left(\rho^p_{ABC}\right)>3,5$& $p>1$&$p>1$& $p >
0.751, 0.861$ \\
 \hline
\multicolumn{1}{|c|}{}&\multicolumn{3}{c|}{}\\${\cal\bar{F}}\left(\rho^p_{ABC}\right)>2,
\bar{{\cal
O}}\left(\rho^p_{ABC}\right)>2$
&
\multicolumn{3}{c|}{
     $p >0.646, 0.772$}  \\[1ex]
\hline
\end{tabular}}
\caption{Theoretical values of the quantum
Fisher information, the observable lower bound defined in Eq.~\ref{lowbound}, and
the conditions witnessing entanglement, Eq.~\ref{witness}, for
the spin observables $J_{3,x(y,z)}$, in $\rho^p_{ABC}$.
The coherence lower bound is an entanglement
witness almost as efficient as the quantum
Fisher information, being blind to
entanglement only for $p\in[0.674,0.751],
[0.646,0.772]$, and to tripartite entanglement
for $p\in [0.813,0.861]$. Note that a more
general sufficient condition for genuine
tripartite entanglement is $|\rho_{1,8}| >
\sqrt{\rho_{2,2} \, \rho_{7,7}} +
\sqrt{\rho_{3,3} \, \rho_{6,6}} +
\sqrt{\rho_{4,4} \, \rho_{5,5}}$, which for
GHZ-diagonal states is also a necessary
condition \cite{entsep}. Hence,
$\rho^p_{ABC}$ is three-partite entangled when
$p > 2^{2/3}-1 \approx 0.587$.}
\label{theory}
\end{table*}
  \begin{figure*} 
 \centering
 \subfigure[]{\includegraphics[width=.3\textwidth]{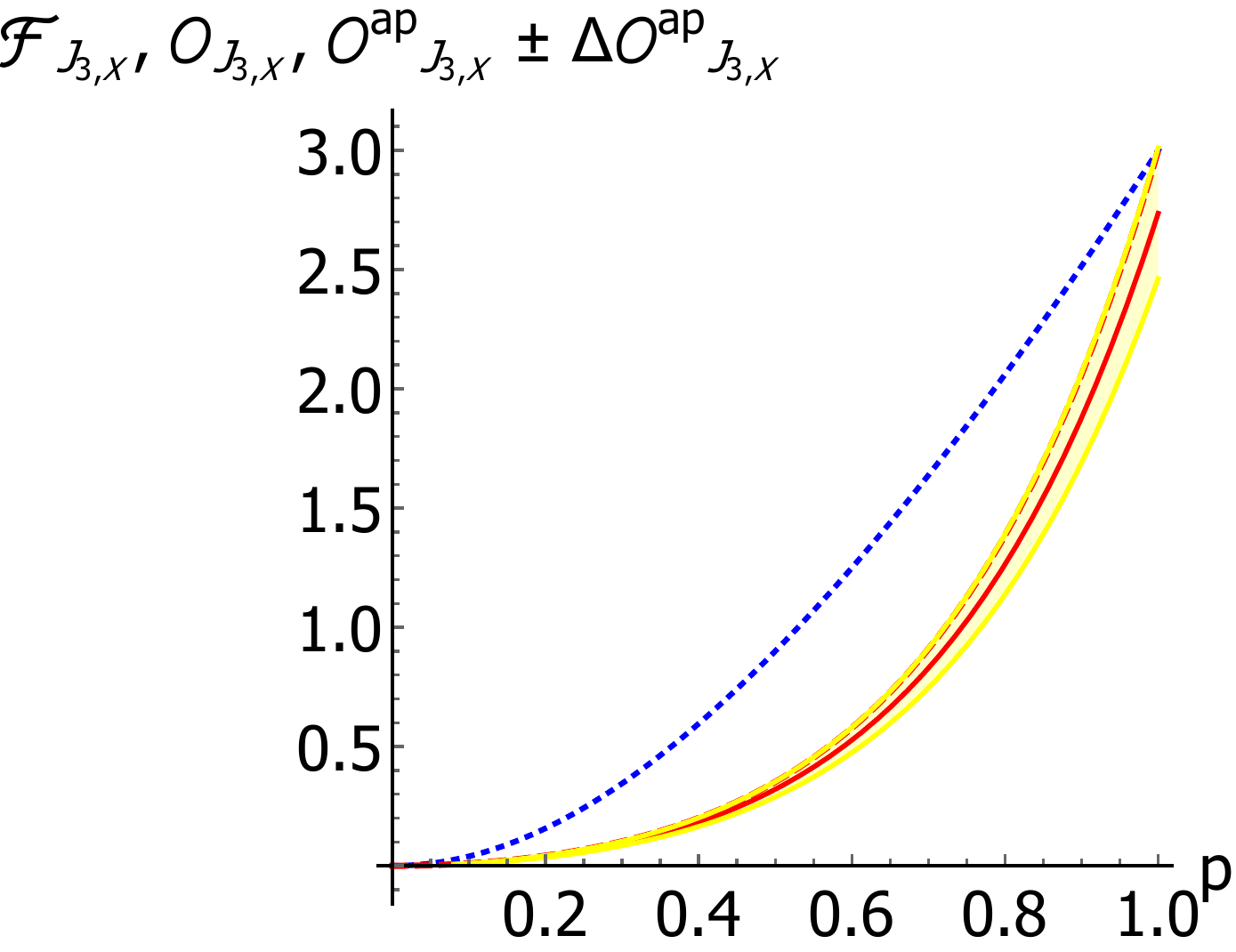}}
\subfigure[]{
\includegraphics[width=.3\textwidth]{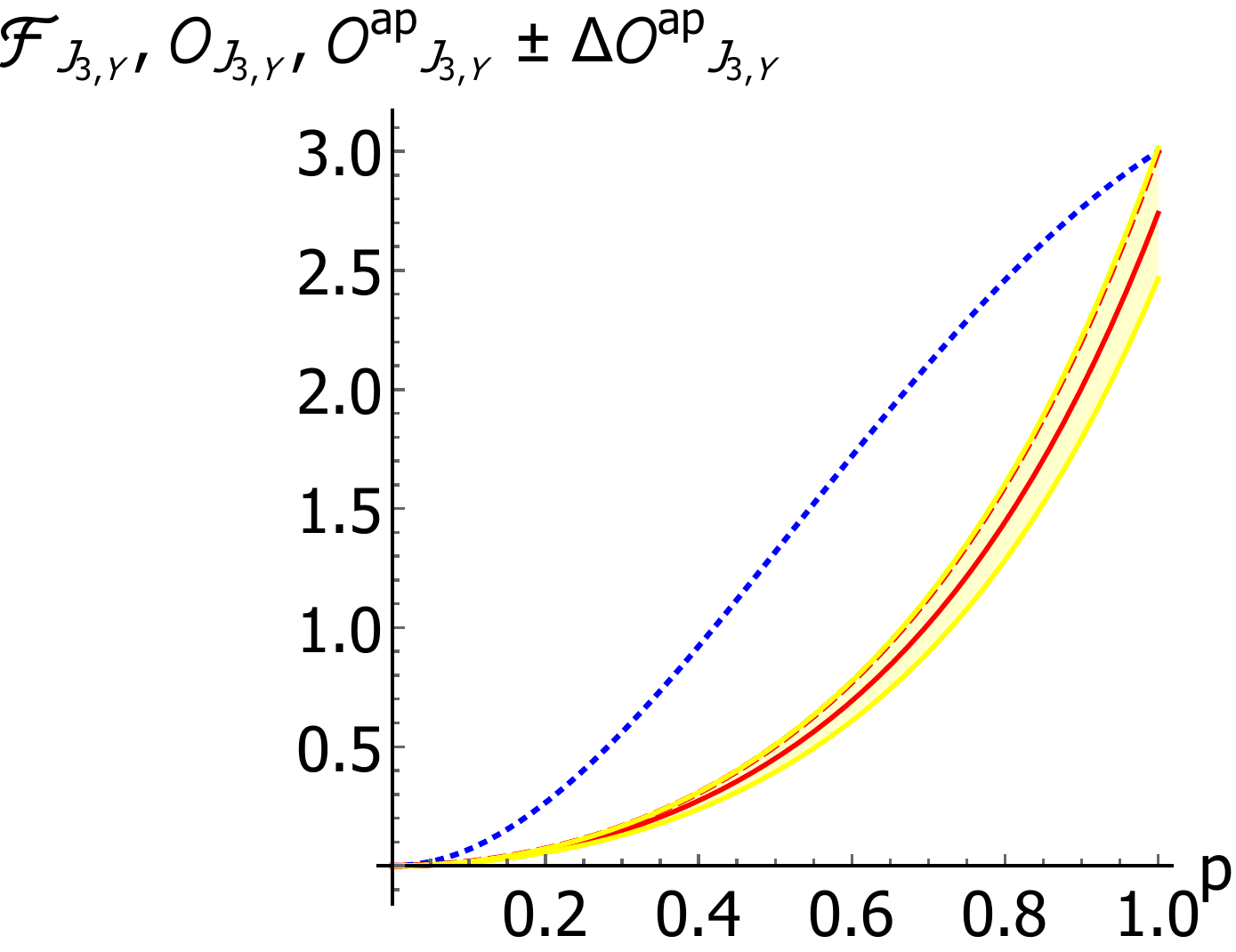}}
\subfigure[]{\includegraphics[width=.3\textwidth]{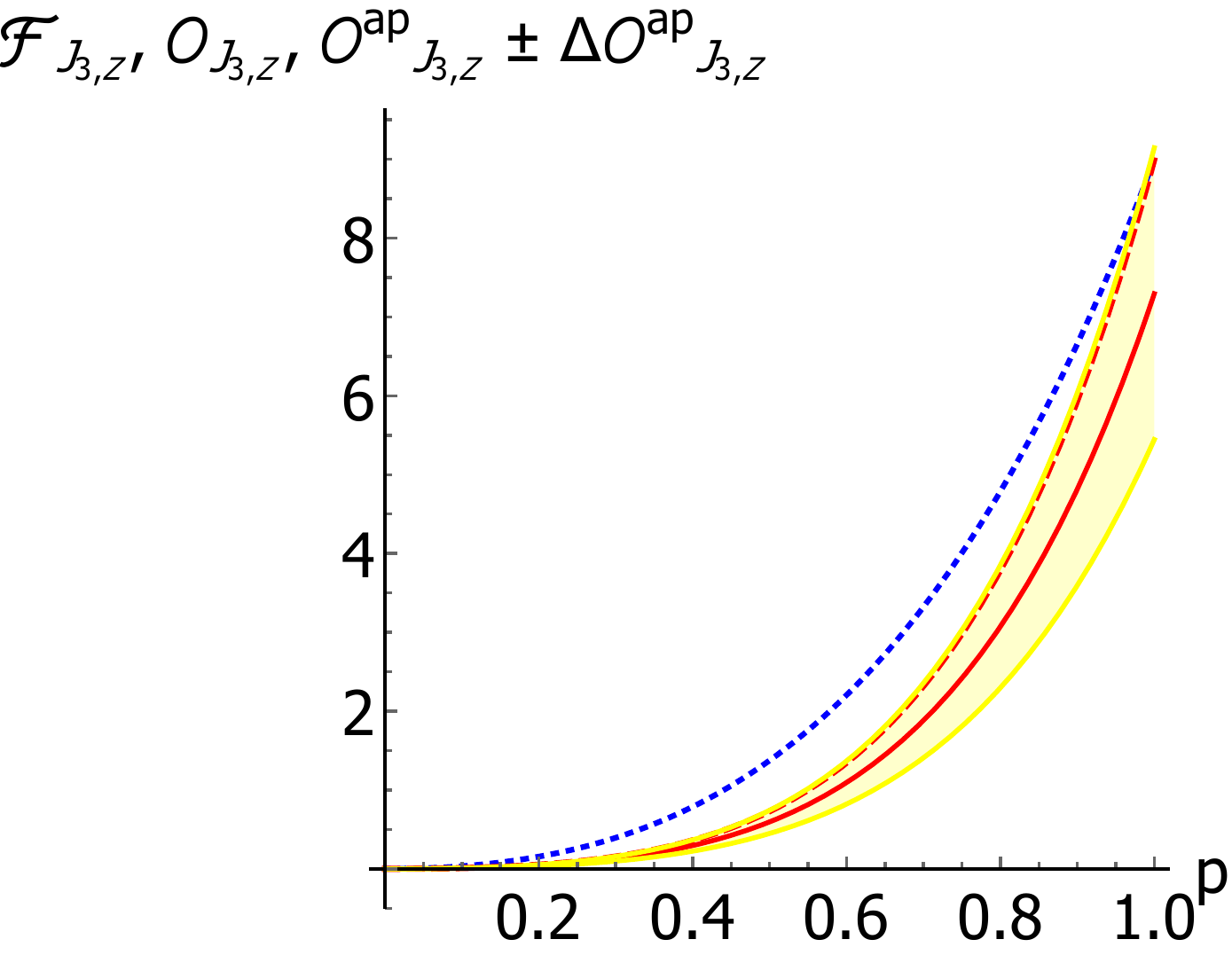}}
\caption{(Colors Online) -- Evaluation of
asymmetry in the state
$\rho^p_{ABC}$ with respect to the
observables $J_{3,x(y,z)}$ (figures (a), (b),
and (c) respectively) as a function of the
mixing parameter $p$. The blue dotted line is
the quantum Fisher information, here showed
for reference, the red dashed line is the
bound ${\cal }O_{J_3}(\rho^p_{ABC})$, the red continuous
line is the approximation ${\cal
}O^{\text{ap}}_{J_3}(\rho^p_{ABC})$ obtained by imposing
$\theta=\pi/6$, and the yellow  band is the error region, whose extreme values are ${\cal
}O^{\text{ap}}_{J_3}(\rho^p_{ABC})\pm \Delta{\cal
}O^{\text{ap}}_{J_3}(\rho^p_{ABC}) $. }
\label{plot}
\subfigure[]{\includegraphics[width=.3\textwidth,height=5cm]{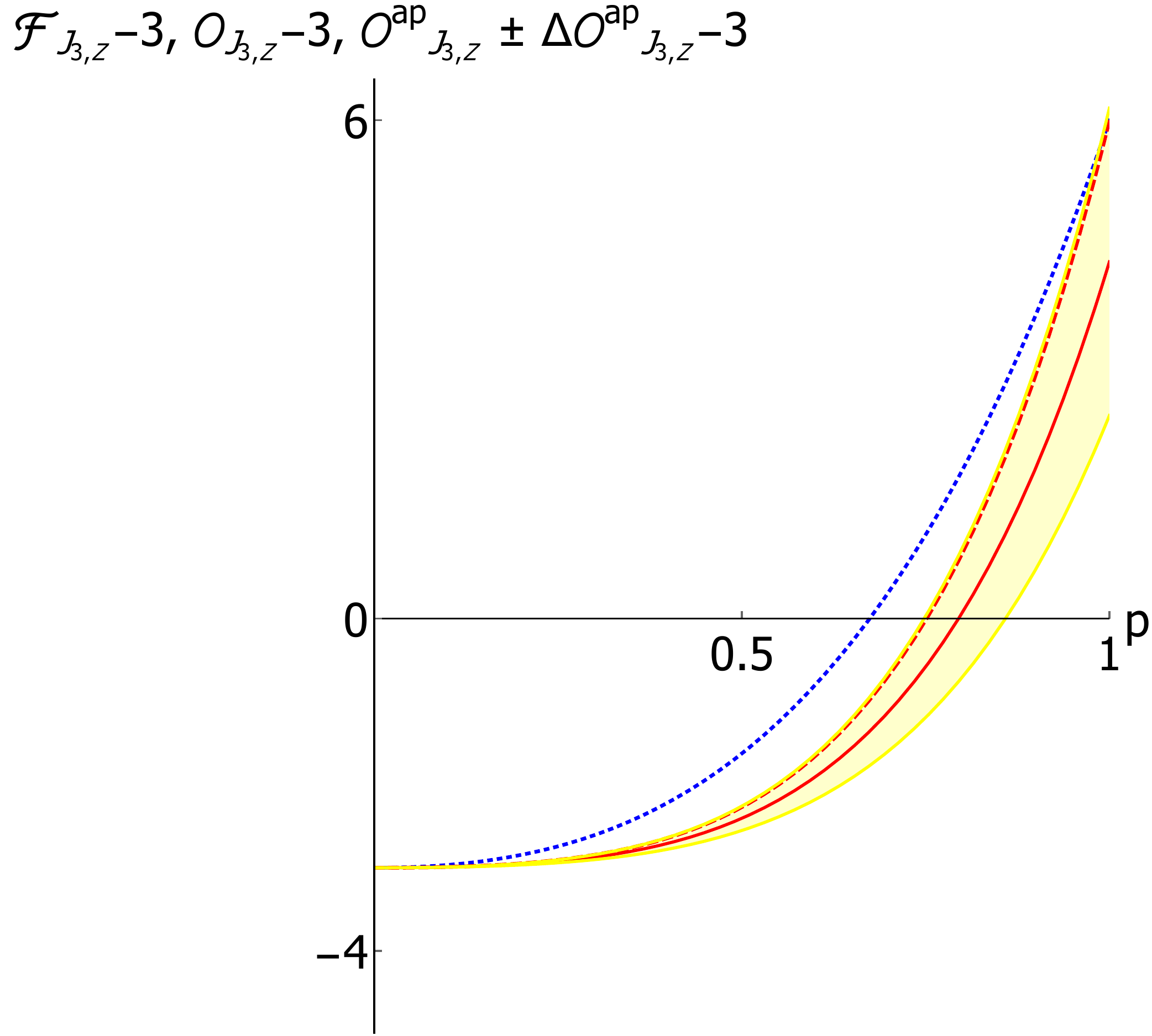}}
\subfigure[]{\includegraphics[width=.3\textwidth,height=5cm]{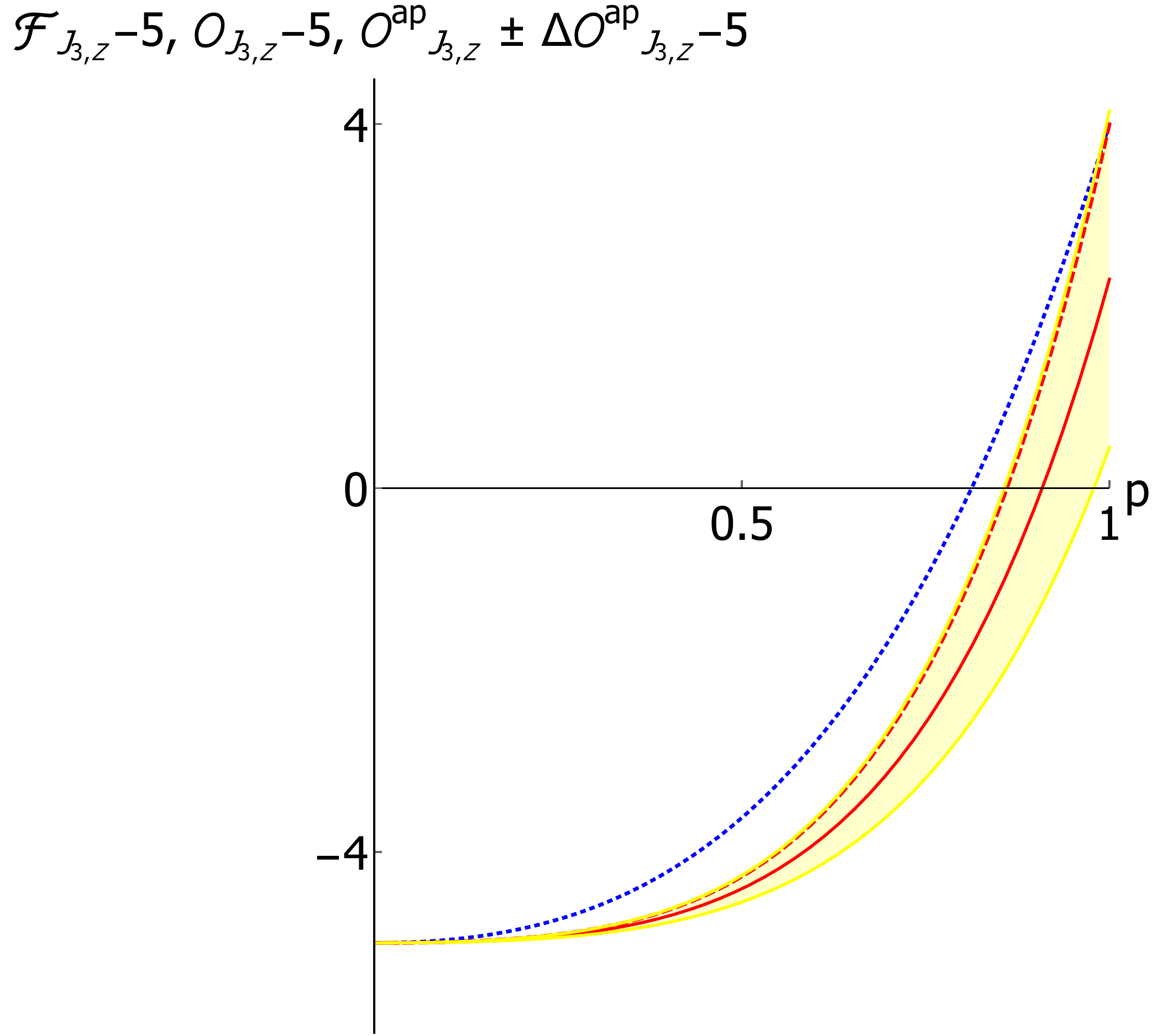}}
\subfigure[]{\includegraphics[width=.3\textwidth,height=5cm]{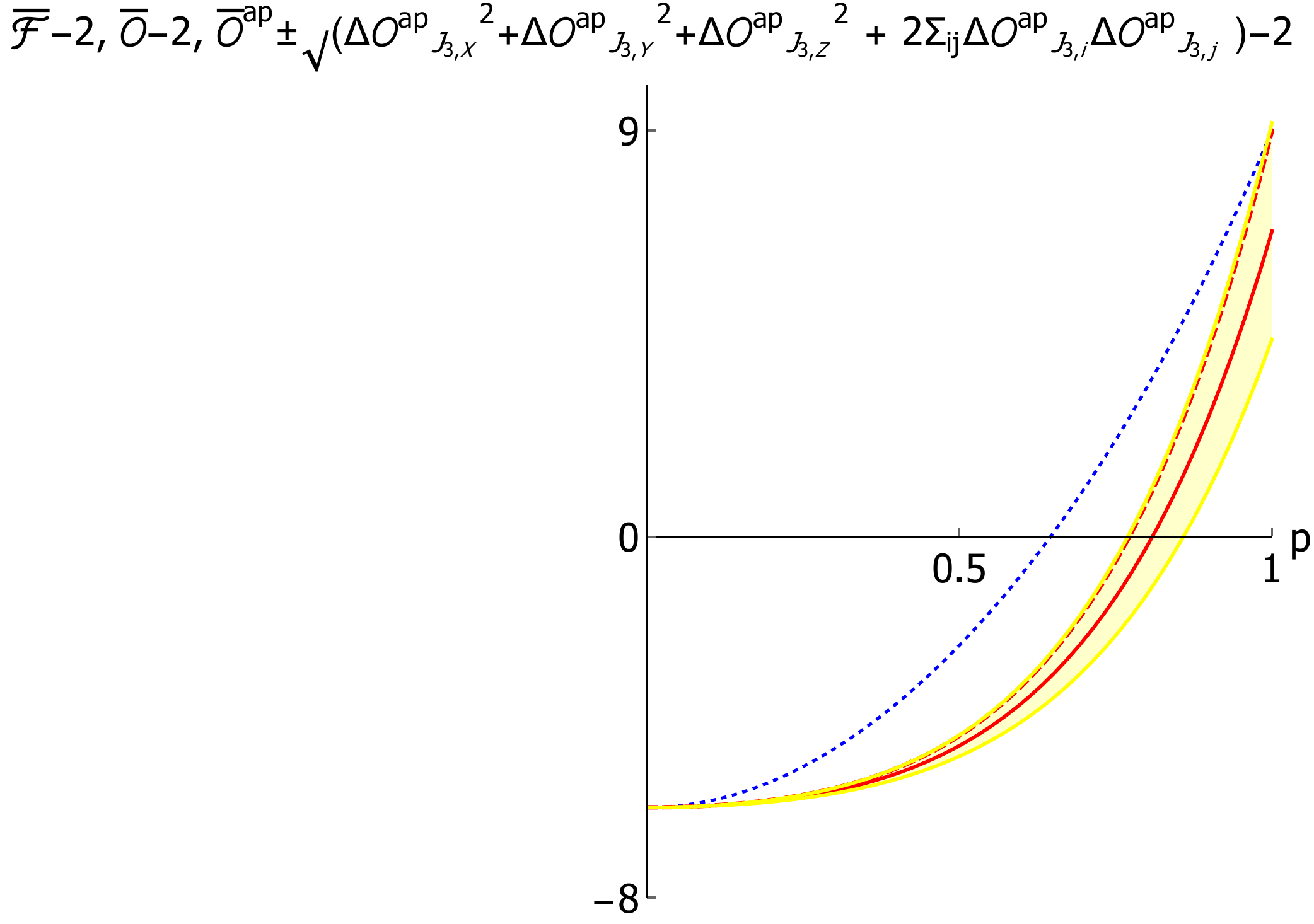}}
\caption{(Colors Online) -- Witnessing entanglement by asymmetry via the inequalities in Eq.~\ref{witness}.  (a) Witnessing 
entanglement in the state $\rho^p_{ABC}$
by computing the quantum Fisher information and the
lower bound, as a function
of the mixing parameter $p$.  The
blue dotted line depicts ${\cal
F}_{J_{z,3}}(\rho^p_{ABC})-3$, the red dashed
line is ${\cal O}_{J_{3,z}}(\rho^p_{ABC})-3$, while the red continuous
line is  ${\cal
}O^{\text{ap}}_{J_3}(\rho^p_{ABC})-3$. Positive values of such
quantities signal entanglement. The yellow band is the error region, bounded by
the extreme values   $({\cal
O}^{\text{ap}}_{J_{3,z}}(\rho^p_{ABC})\pm \Delta {\cal
O}^{\text{ap}}_{J_{3,z}}(\rho^p_{ABC})) -3$. The $J_{3,x(y)}$
cases are not reported as trivially useless,
see Tab.~\ref{theory}. (b) Witnessing  genuine
tripartite entanglement (it is the case $N=3,
k=2$ of Eq.~\ref{witness}). The blue
dotted line depicts ${\cal F}_{J_{3,z}}(\rho^p_{ABC})-5$, the red dashed line
is ${\cal O}_{J_{3,z}}(\rho^p_{ABC})-5$, while the red continuous
line is  ${\cal
}O^{\text{ap}}_{J_3}(\rho^p_{ABC})-5$. 
The error region (yellow) is bounded by the extreme values of $({\cal O}^{\text{ap}}_{J_{3,z}}(\rho^p_{ABC})\pm \Delta {\cal
O}^{\text{ap}}_{J_{3,z}}(\rho^p_{ABC})) -5$. (c) Witnessing entanglement by computing the average values of the quantum Fisher information and the lower bound over a spin basis $\{x,y,z\}$. The blue dotted line is $\bar{{\cal
F}}(\rho^p_{ABC})-2$, the red
dashed line is $\bar{{\cal O}}(\rho^p_{ABC})-2$, while the red continuous
line is  $\bar{{\cal O}}^{\text{ap}}_{J_3} (\rho^p_{ABC})-2$. The yellow 
error region is bounded by
$\left(\bar{{\cal O}}^{\text{ap}}(\rho^p_{ABC})\pm
\sqrt{\sum_i{\Delta {\cal
O}^{\text{ap}}_{J_{3,i}}}^2(\rho^p_{ABC})+2\sum_{ij}
\Delta {\cal O}^{\text{ap}}_{J_{3,i}} \Delta {\cal
O}^{\text{ap}}_{J_{3,i}}}\right) -5$.}
\label{wit}
\end{figure*} 

\subsubsection*{Closed formula for the asymmetry lower bound in multi-qubit systems}
We here provide a closed formula for 
the asymmetry lower bound in one, two and three-qubit states, with respect to additive  Hamiltonians $H_N=\sum_{i=1}^N h_i,
h_i=I_{1,2,\ldots,i-1}\otimes
h_i\otimes I_{i+1,i+2,\ldots,N},
h_i=1/2 \sigma_{i}$,  
$\sigma_{i}$ representing spin-1/2 observables, e.g. the Pauli matrices.
By recalling that $e^{i \sigma 
\theta/2}= \cos{\theta/2}~I_2+ i
\sin{\theta/2}~\sigma $, we get an exact
expression for the  lower bound in terms of phase shifts
$U_{H_N}(\theta)=e^{-i H_N \theta}$. For $N=1, H_1=h$,
one has
\begin{eqnarray}\label{one}
{\cal O}_{H_1}(\rho)&=&\text{Tr}[\rho^2]-\text{Tr}[\rho
U_h(\pi) \rho U_h^{\dagger}(\pi)],  
\end{eqnarray}
For  $N=2, H_2=h_1+h_2$:  
\begin{eqnarray}\label{two}
{\cal
O}_{H_2}(\rho)&=&3\text{Tr}[\rho^2]-4\text{Tr}[\rho
U_{H_2}(\pi/2) \rho
U_{H_2}^{\dagger}(\pi/2)]\nonumber\\
&+&\text{Tr}[\rho U_{H_2}(\pi) \rho
U_{H_2}^{\dagger}(\pi)], \nonumber\\
U_{H_2}(\theta)&=&U_{h_1}(\theta)U_{h_2}(\theta). 
\end{eqnarray}
 For $N=3, H_3=h_1+h_2+h_3$:
\begin{eqnarray}\label{three}{\cal O}_{H_3}(\rho)
&=&6\text{Tr}[\rho^2]-4\Big\{\text{Tr}[\rho
U_{h_1+h_2}(\pi/2)\rho
U_{h_1+h_2}^{\dagger}(\pi/2)] \nonumber\\
&+& \text{Tr}[\rho U_{h_1+h_3}(\pi/2)\rho
U_{h_1+h_3}^{\dagger}(\pi/2)]\nonumber\\
&+&\text{Tr}[\rho U_{h_2+h_3}(\pi/2)\rho
U_{h_2+h_3}^{\dagger}(\pi/2)]\Big\}\nonumber\\
&+&\text{Tr}[\rho U_{h_1+h_2}(\pi)\rho
U_{h_1+h_2}^{\dagger}(\pi)]+\text{Tr}[\rho
U_{h_1+h_3}(\pi)\rho
U_{h_1+h_3}^{\dagger}(\pi)]\nonumber\\
&+&\text{Tr}[\rho U_{h_2+h_3}(\pi)\rho
U_{h_2+h_3}^{\dagger}(\pi)]\nonumber\\
&+&\text{Tr}[\rho U_{h_1}(\pi)\rho
U_{h_1}^{\dagger}(\pi)]+\text{Tr}[\rho
U_{h_2}(\pi)\rho
U_{h_2}^{\dagger}(\pi)]\nonumber\\
&+&\text{Tr}[\rho U_{h_3}(\pi)\rho
U_{h_3}^{\dagger}(\pi)].
\end{eqnarray}
We conjecture that it is possible to iterate the procedure and work out equivalent expressions for an arbitrary number of qubits. 
\section{Detection of multipartite entanglement via asymmetry}
\subsection{Asymmetry witnesses Entanglement}
\label{sect3a}

It is often desirable to consider a high
dimensional system as a partition of
subsystems. Such a partition is usually
dictated by the physical constraints of the
problem, for example the spatial separation
between the parts of the system. It is then
interesting to understand the interplay
between asymmetry with respect to a global
observable and the quantum properties of the
subsystems. In spite of being a
basis-dependent feature,  coherence is
linked to basis-independent features of
multipartite systems as quantum correlations
\cite{lqu,ger2,china,alta,morimae}. Here we
show that, for an $N$-qubit system, the
observable asymmetry bound ${\cal O}_H(\rho)$
measured on the global system state witnesses
entanglement between the partitions. There are
several entanglement witnesses written in
terms of the quantum Fisher information. They
relate entanglement to the system speed of response to phase shifts generated by additive
spin-$1/2$ Hamiltonians $J_{N}=\sum_{i=1}^N
1/2\sigma_i$
\cite{toth,toth2,toth4,smerzi,smerzi2}. In
particular,
a constraint which cannot be satisfied by
$k$-separable states of $N$ qubits is ${\cal
F}_{J_N}(\rho) \geq nk^2 + (N-nk)^2 $,
where $n = \lfloor \frac{N}{k} \rfloor$.
Thus, verifying this relation certifies genuine $k$-partite
entanglement \cite{horo}. Also, if $\bar{{\cal
F}}(\rho)=1/3({\cal F}_{J_{N,x}}(\rho)+{\cal
F}_{J_{N,y}}(\rho)+{\cal F}_{J_{N,z}}(\rho))>2N/3$,
then the state is entangled. Therefore, if
there exists a spin basis $\{x,y,z\}$ such that 
 the following conditions are
satisfied:\begin{eqnarray}\label{witness}
{\cal O}_{J_{N,x(y,z)}}(\rho) &>& nk^2 +
(N-nk)^2,\\
\bar{\cal O}(\rho)&=&1/3({\cal
O}_{J_{N,x}}(\rho)+{\cal O}_{J_{N,y}}(\rho)+{\cal
O}_{J_{N,z}}(\rho))> 2N/3,\nonumber
\end{eqnarray}
a state $\rho$ is respectively genuinely
$k$-partite entangled and entangled.

  \subsection{A case
study}\label{sect3b}
Here we apply our scheme to simulate the non-tomographic
detection of asymmetry and entanglement in a
three-qubit state.
We choose as probe state the GHZ-diagonal
state $\rho^p_{\text{ABC}}$. This allows one to investigate the
behavior of the asymmetry lower bound and
entanglement witness in the presence of noise
in the system. The two copies of the GHZ
diagonal state $\rho^p_{A_1B_1C_1},
\rho^p_{A_2B_2C_2}$ are obtained by
initializing a six qubit processor in
$\rho_i=1/2(I_2+p\sigma_z),
i=i\ldots,6$, and applying Hadamard and CNOT
gates as described in Fig.~\ref{complete}.\\
We measure the asymmetry of the input state
with respect to the set of spin Hamiltonians
$J_{3}=\sum_{i=A,B,C} j_{3,i},
j_{3,A}=j_A\otimes I_{BC},j_{3,B}=I_{A}\otimes
j_B\otimes I_{C},j_{3,C}=I_{AB}\otimes
j_C , j =1/2 \sigma_{x,(y,z)},$ by
computing the values of the lower bound, and the approximation defined in
Eq.~\ref{app}, for each observable. Of course,
we may obtain the asymmetry with respect to
any self-adjoint operator in the three-qubit
Hilbert space.
This is done by implementing the unitary gate
$U_{J_3}(\theta)=U_{j,A}(\theta)\otimes
U_{j,B}(\theta)\otimes U_{j,C}(\theta)$ on
a copy of the state and then   building up an interferometric
configuration (Fig.~\ref{complete}). Performing the
polarisation measurements on the ancillary
qubit makes possible to determine ${\cal O}_{J_{3}}(\rho^p_{ABC})$. We select a small but experimentally plausible phase shift,  $ \theta=\pi/6$ \cite{speed}. Obviously, to
evaluate the purity, no gate has to be applied. The purity and overlap values extracted by the quantities
$\text{Tr}[\rho^p_{ABC}
U_{J_3}(\pi/6)\rho^p_{ABC}U_{J_3}^{\dagger}(\pi/6)]$
determine ${\cal
O}_{J_3}^{\text{ap}}(\rho^p_{ABC})$.
No further action is necessary to verify the
presence of entanglement through the witnesses
in Eq.~\ref{witness}, as the values of ${\cal
O}^{\text {ap}}_{J_3}(\rho^p_{ABC})$ have been obtained in the
previous steps. For $N=3$, we have $k=1
\Rightarrow {\cal O}_{J_3}(\rho^p_{ABC}) \geq 3, k=2
\Rightarrow {\cal O}_{J_3}(\rho^p_{ABC}) \geq 5,$ and
${\cal\bar{O}}(\rho^p_{ABC})>2$. The results are summarised
in Tab.~\ref{theory} and Figs.~\ref{plot},\ref{wit}.
 
\section{Conclusion}\label{sect4} 
In this work, we provided an experimental
recipe to witness multipartite entanglement by detecting
asymmetry with respect to an additive Hamiltonian. We employed
an experimentally friendly lower bound of the
quantum Fisher information to quantify
asymmetry, a geometric property of quantum
systems underpinned by coherence in an
observable eigenbasis. The scheme is suitable
for detection of asymmetry in large scale
quantum registers, as it requires a limited
number of measurements regardless the
dimension of the system.  We
showed that in multipartite states the
asymmetry lower bound with respect to additive
observables is a witness of multipartite
entanglement. Our results suggest further
lines of investigation. To the best of our
knowledge, the lower bound ${\cal O}_H$ is the first faithful experimental
quantifier of asymmetry for finite dimensional
systems. Thus, on the experimental side, we
call for a demonstration of our study. Moreover, we observe that a quadratic
($O(N^2)$) sensitivity to phase shifts
generated by additive Hamiltonian in $N$-party
systems, as measured by the quantum Fisher
information, has been associated to another
elusive quantum effect, i.e. quantum
macroscopicity \cite{leggett,frow,ben}. It is
  clear that high values of coherence are
essential to quantum macroscopicity, yet the
interplay between the two concepts still needs
to be clarified.

\section*{Acknowledgments}
We thank Hang Li and Geza T\'{o}th for
fruitful discussions. This work was supported
by the EPSRC (UK) and the Wolfson College,
University of Oxford.


\begin{thebibliography}{99}
 
\bibitem{opp}M. Horodecki and J. Oppenheim,
{\it Int. J. Mod. Phys. B} {\bf 27}, 1345019
(2013).
%%%%

\bibitem{superreview}S. D. Bartlett, T.
Rudolph, and R. W. Spekkens, {\it Rev. Mod.
Phys.} {\bf 79}, 555 (2007).

\bibitem{gour}G. Gour and R.W. Spekkens, {\it
New J. Phys.} {\bf 10}, 033023 (2008).

\bibitem{newmar}I. Marvian, {\it Symmetry,
Asymmetry and Quantum Information}, Phd
Thesis, University of Waterloo (2012).

\bibitem{mehdi}M. Ahmadi, D. Jennings, and T.
Rudolph, {\it New J. Phys.} {\bf 15}, 013057
(2013).

\bibitem{lqu}D. Girolami, T., Tufarelli, and
G. Adesso, {\it Phys. Rev. Lett.} {\bf 110},
240402 (2013).

 

\bibitem{speknat}I. Marvian and R. W.
Spekkens, {\it Nature Comm.} {\bf 5}, 3821
(2014).

\bibitem{me}D. Girolami, {\it Phys. Rev.
Lett.} {\bf 113}, 170401 (2014).
 
\bibitem{blind}D. Girolami {\it et al.}, {\it
Phys. Rev. Lett} {\bf 112}, 210401 (2014).
%%%%112 (21), 210401

\bibitem{luo}S. Luo, {\it Phys. Rev. Lett.}
{\bf 91}, 180403 (2003).

\bibitem{aberg}J. Aberg, {\it Phys. Rev.
Lett.} {\bf 113}, 150402 (2014).

\bibitem{luocri}S. Luo, Theor. Math. Phys.
{\bf 143}, 681 (2005).

\bibitem{luo2}S. Luo, S. Fu, and C. H. Oh,
{\it Phys. Rev. A} {\bf 85}, 032117 (2012).

\bibitem{herbut}F. Herbut, {\it J. of Phys. A}
{\bf 38}, 2959 (2005).

\bibitem{plenio}T. Baumgratz, M. Cramer, and
M. B. Plenio, {\it Phys. Rev. Lett.} {\bf
113}, 140401 (2014).

\bibitem{china}Y. Yao, X. Xiao, L. Ge, and C.
P. Sun, {\it Phys. Rev. A} {\bf 92}, 022112
(2015).

\bibitem{china2}S. Du, Z. Bai, and Y, Guo,
{\it Phys. Rev. A} {\bf 91}, 052120 (2015).

\bibitem{ger2}A. Streltsov, U. Singh, H. S.
Dhar, M. N. Bera, and G. Adesso, {\it Phys.
Rev. Lett.} {\bf 115}, 020403 (2015).

\bibitem{heng}Z. Xi, Y. Li, and H. Fan, {\it
Sci. Rep.} {\bf 5}, 10922 (2015).

%%%%%

\bibitem{noriwit}C.-M. Li, N. Lambert, Y.-N.
Chen, G.-Y. Chen, and F. Nori, {\it Sci. Rep.}
{\bf 2}, 885 (2012).

\bibitem{agata}A. Monras, A. Checinska, and A.
K. Ekert, {\it New J. Phys.} {\bf 16}, 063041
(2014).

\bibitem{felix}F. A. Pollock, A. Checinska, S.
Pascazio, and K. Modi, arXiv:1507.05051.


\bibitem{alta}C. Altafini, {\it Phys. Rev. A}
{\bf 012311} (2004).

\bibitem{ger1}T. R. Bromley, M. Cianciaruso,
and G. Adesso, {\it Phys. Rev. Lett.} {\bf
114}, 210401 (2015).


%%%%
\bibitem{horo}R. Horodecki, P. Horodecki, M.
Horodecki, and K. Horodecki, {\it Rev. Mod.
Phys.} {\bf 81}, 865 (2009).


 


\bibitem{jozsa}R. Jozsa and N. Linden, {\it
Proc. R. Soc. Lon A.} {\bf 459}, 2011 (2003).

\bibitem{metrorev}V. Giovannetti, S. Lloyd,
and L. Maccone, {\it Nature Photon.} {\bf 5},
222 (2011).

\bibitem{morimae}A. Shimizu and T. Morimae,
{\it Phys. Rev. Lett.} {\bf 95}, 090401
(2005).

\bibitem{huber}M. Huber, F. Mintert, A.
Gabriel and B. C. Hiesmayer, {\it Phys. Rev.
Lett.} {\bf 104}, 210501 (2010).


\bibitem{toth3}O. G\"{u}hne and G. T\'{o}th,
{\it Phys. Rep.} {\bf 474}, 1 (2009).

\bibitem{helstrom}C. W. Helstrom, {\it Quantum
detection and estimation theory} (Academic
Press, New York, 1976).


\bibitem{toth}G. T\'{o}th and I. Apellaniz,
{\it J. Phys. A: Math. Theor.} {\bf 47},
424006 (2014).


\bibitem{speed}C. Zhang, {\it et al.},
arXiv:1611.02004.
%%%%% 

\bibitem{paz}J. P. Paz and A. Roncaglia, {\it
Phys. Rev. A} {\bf 68}, 052316 (2003).

\bibitem{brun}T. A. Brun, {\it Quant. Inf. and
Comp.} {\bf 4}, 401 (2004).
 
\bibitem{dariano}G. M. D'Ariano and P.
Perinotti, {\it Phys. Rev. Lett.} {\bf 94},
090401 (2005).

\bibitem{ekertdir}A. K. Ekert, C. Moura Alves,
D. K. L. Oi, M. Horodecki, P. Horodecki, and
L. C. Kwek, {\it Phys. Rev. Lett.} {\bf 88},
217901 (2002).

\bibitem{filip}R. Filip, {\it Phys. Rev. A}
{\bf 65}, 062320 (2002).

%%%%%%%%%



\bibitem{jeong2}H. Jeong, C. Noh, S. Bae, D.
G. Angelakis, and T. C. Ralph, {\it J. Opt.
Soc. Am. B} {\bf 31}, 3057 (2014).
 
\bibitem{pasca}H. Nakazato, T. Tanaka, K.
Yuasa, G. Florio, and S. Pascazio, {\it Phys.
Rev. A } {\bf 85}, 042316 (2012).
 


\bibitem{mintert}F. Mintert and A.
Buchleitner, {\it Phys. Rev. Lett.} {\bf 98},
140505 (2007).

\bibitem{mintert2}S. P. Walborn, P. H. Souto
Ribeiro, L. Davidovich, F. Mintert, and A.
Buchleitner, {\it Nature} {\bf 440}, 1022
(2006).

\bibitem{kus}M. Oszmaniec and M. Kuś
{\it Phys. Rev. A} {\bf 88}, 052328 (2013).

\bibitem{geodiscordchina}J. Jin, F. Zhang, C.
Yu, and H. Song
{\it J. Phys. A: Math. Theor.} {\bf 45},
115308 (2012).
  


\bibitem{yu}S. Yu, arXiv:1302.5311. 

\bibitem{toth2}G. T\'{o}th and D. Petz, {\it
Phys. Rev. A} {\bf 87}, 032324 (2013).

 \bibitem{petzmono}D. Petz, {\it Lin. Alg. Appl.} {\bf 244,} 81 (1996).

\bibitem{gibilisco}P. Gibilisco, D. Imparato, and T. Isola, {\it Proc. of the Am. Math. Soc.} {\bf 137}, 317 (2008).

\bibitem{diogo}D. Paiva Pires, L. C.
C\'{e}leri, and D. O. Soares-Pinto, {\it Phys.
Rev. A} {\bf 91}, 042330 (2015).

%\bibitem{caves}S. Braunstein, C. M. Caves,{\it Phys. Rev. Lett} {\bf 72}, 3439 (1994).
%\bibitem{geo}I. Bengtsson and K. Zyczkowski,{\it Geometry of Quantum States} (CambridgeUniversity Press, Cambridge, 2007).


\bibitem{schirmer}X. Wang and S. G. Schirmer,
{\it Phys. Rev. A} {\bf 79}, 052326 (2009).
 
\bibitem{tufo}T. Tufarelli, D. Girolami, R.
Vasile, S. Bose, and G. Adesso, {Phys. Rev. A}
{\bf 86}, 052326 (2012).

\bibitem{piani} M. Piani, {\it Phys. Rev. A}
{\bf 86}, 034101 (2012).
%%%%%%%%%%%%%

\bibitem{rugg}D. Girolami, R. Vasile, and G.
Adesso, {\it Int. J. of Mod. Phys. B} {\bf
27}, 1345020 (2013).

 


%%%%%%%%%

 
\bibitem{smerzi}L. Pezz\'{e} and A. Smerzi,
{\it Phys. Rev. Lett.} {\bf 110}, 163604
(2013).

\bibitem{toth4}G. T\'{o}th, {\it Phys. Rev. A}
{\bf 85}, 022322 (2012).

\bibitem{smerzi2}P. Hyllus, W. Laskowski, R.
Krischek, C. Schwemmer, W. Wieczorek, H.
Weinfurter, L. Pezz\'{e}, and A.
Smerzi, {\it Phys. Rev. A} {\bf 85}, 022321
(2012).

\bibitem{entsep}O. G\"{u}hne and M. Seevinck,
{\it New J. Phys.} {\bf 12}, 053002 (2010).
 
\bibitem{leggett}A. Leggett, {\it Prog. Theor.
Phys. Supp.} {\bf 69}, 80 (1980).
 
    
\bibitem{frow}F. Fr\"{o}wis and W. D\"{u}r,
{\it New J. Phys.} {\bf 14}, 093039 (2012).
 
\bibitem{ben}B. Yadin and V. Vedral, {\it
Phys. Rev. A} {\bf 93}, 022122 (2016).

 
%%%%%%%%%%%%%%%
 


  \end{thebibliography}
    \end{document}